\documentclass{article}

\usepackage{epsfig,graphicx}
\usepackage{epstopdf}

\begin{document}

\begin{titlepage}

\title{Comparison of the full matrix element calculation with the
simplified calculation of the $W_R$ production at the LHC (left-right
symmetric model)}

\author{G.~Bednik \footnote{INR (Moscow), MSU} \ \ \
        M.~Kirsanov \footnote{INR (Moscow), e-mail: Mikhail.Kirsanov@cern.ch} }

\end{titlepage}

\maketitle

\abstract{We estimate possible systematic error that can arise in the
simplified simulation of the $W_R$ production and its subsequent decay
into heavy neutrino at LHC (left-right symmetric model). The general-purpose
event generators simulate this process as a $W_R$ production followed by
two decays, loosing full information about the polarization of the intermediate
particles. We calculate the cross section using the full matrix elements
with propagators, perform the simulation and compare the results with
the simplified simulation.}

\clearpage

\section{Introduction}
It is known that in the Standard Model there are left-handed W-bosons that
interact with left-handed fermions. The Left-right symmetric model (for more
details see \cite{Mohapatra}) states that there are also right-handed W bosons
interacting with right handed-fermions. In the minimal Left-right symmetric
model they have the same coupling constants to the fermions but are much
heavyer, otherwise they would be already discovered. In other words,
in the SM the terms responsible for the interaction between fermions and
weak bosons are
\begin{eqnarray}
i \bar{Q_L} \gamma^{\mu} \left(  \partial_{\mu} - \frac{ig}{2}\vec \tau \vec (W_L)_{\mu} - \frac{ig'}{6} B_{\mu} \right) Q_L + \nonumber\\
i \bar{\psi_L} \gamma^{\mu} \left(  \partial_{\mu} - \frac{ig}{2}\vec \tau \vec (W_L)_{\mu} + \frac{ig'}{2} B_{\mu} \right) \psi_L .
\end{eqnarray} 
Here Q are quark doublets and $\psi$ are lepton doublets.
The Left-right symmetric model contains also the following terms:
\begin{eqnarray}
i \bar{Q_R} \gamma^{\mu} \left(  \partial_{\mu} - \frac{ig}{2}\vec \tau \vec (W_R)_{\mu} - \frac{ig'}{6} B_{\mu} \right) Q_R + \nonumber\\
i \bar{\psi_R} \gamma^{\mu} \left(  \partial_{\mu} - \frac{ig}{2}\vec \tau \vec (W_R)_{\mu} + \frac{ig'}{2} B_{\mu} \right) \psi_R .
\end{eqnarray} 
 The lepton doublets here contain ordinary charged leptons and heavy
right-handed neutrinos. We assume that they can be Dirac or Majorana
neutrinos.

The searches for these new particles are being performed now at LHC \cite{CMS}.
The purpose of this work is to estimate possible systematic errors arising
from some approximations usually used in the simulation of the signal
by general purpose event generators.

\begin{figure}[h!]
\begin{center}
\includegraphics[width=16cm,angle=0]{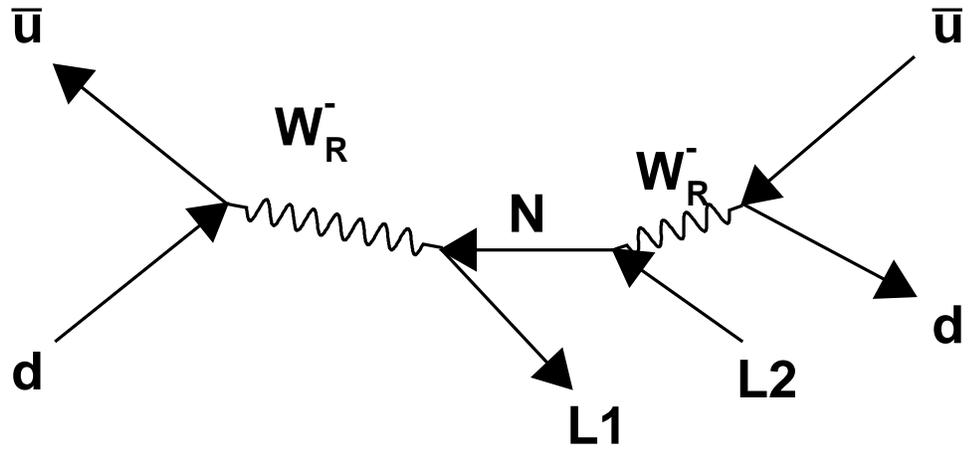}
\caption{Diagram with Dirac neutrino}
\label{DiracDiagram}
\end{center}
\end{figure}

We consider the following process. $d$-quark and $u$-antiquark interact with
each other producing the $W^-_R$ -boson. It decays into the lepton $L1$ and
heavy antineutrino $N$ that decays into another lepton $L2$, $u$-antiquark and
$d$-quark via the virtual $W^-_R$-boson. This process is represented in
Figure \ref{DiracDiagram}.


\begin{figure}[h!]
\begin{center}
\includegraphics[width=16cm,angle=0]{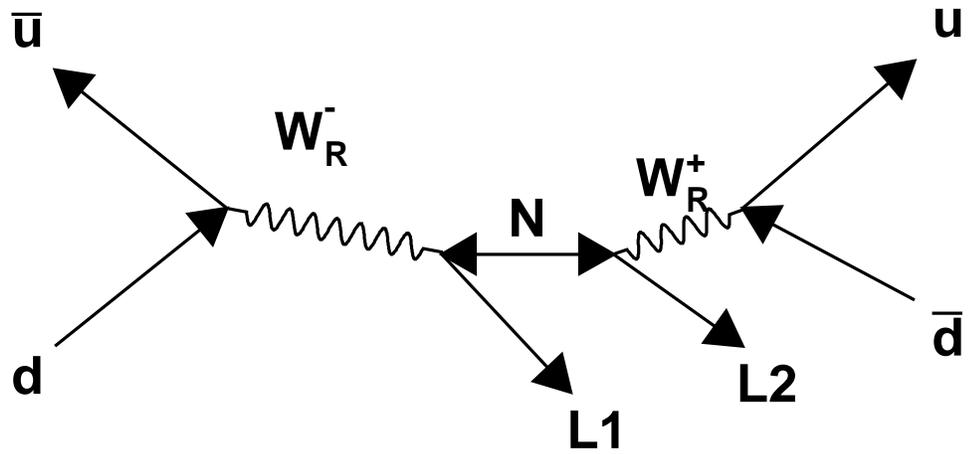}
\caption{Diagram possible with Majorana heavy neutrino (leptons have same sign)}
\label{MajoranaDiagram}
\end{center}
\end{figure}

This process takes place regardless of the type of the neutrino: it can be
either Dirac-like or Majorana-like. However if the neutrino is Majorana-like,
another process is possible with the same initial particles:
$ d \bar u \to L_1 L_2 u \bar d $ (it is drawn in Figure \ref{MajoranaDiagram}).


The cross-section of these reactions can be calculated explicitly.
However, the 4 - particle final state is too complicated for the general
purpose event generators (for example pythia6 \cite{Sjostrand:2006za})
usually used in the HEP analyses. In such generators this process is considered
in the following approximate way: quark-antiquark pair produces real
$W^-$-boson, then it decays into real lepton $L2$ and antineutrino and
the antineutrino decays into antilepton $L2$  and quark-antiquark pair
(if the neutrino is Majorana-like it can decay either to lepton or to
antilepton). In this case the information about the polarization of intermediate
particles is lost. We consider both approaches and compare the results they
give. Let us note that we neglect other possible reactions with the same
initial and final state particles.

\rule{0pt}{10mm}

\section{Cross section with the full matrix element}
\label{Cross-section}

Let the particles in our process have the following momenta:
\begin{itemize}
\item[$p_1$] -momentum of the initial $d$-quark
\item[$p_2$] -momentum of the initial $u$-antiquark
\item[$k_1$] -momentum of the lepton $L1$
\item[$k_2$] -momentum of the lepton $L2$
\item[$q_1$] -momentum of the final $d$-quark
\item[$q_2$] -momentum of the final $u$-antiquark
\end{itemize}
The amplitude of the reaction represented in Figure \ref{DiracDiagram} is
written in the following way (here we use the common notations that can be
found for example in \cite{Peskin}, \cite{CL}):
\begin{eqnarray}
\bar{v}(p_2) \frac{ig}{\sqrt{2}} \gamma^{\mu} \frac{(1+\gamma^5)}{2} u(p_1) \frac{-i}{(p_1+p_2)^2 - M_W^2} \left( g_{\mu\nu} \frac{(p_1+p_2)_{\mu}(p_1+p_2)_{\nu}}{M_W^2} \right) \nonumber\\
\bar{u}(k_1) \frac{ig}{\sqrt{2}} \gamma^{\nu} \frac{(1+\gamma^5)}{2} \frac{i(\hat p_1 + \hat p_2 - \hat k_1) + m_{\nu}}{(p_1+p_2-k_1)^2 - m_{\nu}^2} \frac{ig}{\sqrt{2}} \gamma^{\lambda} \frac{(1+\gamma^5)}{2} v(k_2) \nonumber\\
\frac{-i}{(q_1+q_2)^2 - M_W^2}  \left( g_{\lambda \rho} \frac{(q_1+q_2)_{\lambda}(q_1+q_2)_{\rho}}{M_W^2} \right) \frac{ig}{\sqrt{2}} \bar{u} (q_1) \gamma^{\rho} \frac{(1+\gamma^5)}{2} v(q_2)
\end{eqnarray}
Here we do not take into account any possible mixings among the particles. Further we also neglect the masses of all particles except $W^-$boson and the neutrino. Using the standard rules (see \cite{Peskin}) one can show that
\begin{eqnarray}
|M|^2 = \frac{4 g^8 (k_1 p_2) (q_2 k_2) ((k_1 p_2)(p_1 q_1) + (p_1(k_1-p_2))(q_1(k_1-p_2))       ) }{ ((p_1+p_2)^2-M_W^2)^2 ((q_1+q_2)^2-M_W^2) ((p_1+p_2-k_1)^2 - m_{\nu}^2)     }
\end{eqnarray}
We used the \textsc{mathematica} package \textsc{feyncalc} to obtain this expression.

Then this formula is substituted into the general expression for cross-section of massless particles \cite{LL}
\begin{eqnarray}
d \sigma = \frac{1}{(2\pi)^8} \delta^{(4)} (k_1+k_2+q_1+q_2-p_1-p_2) |M|^2 \frac{1}{4(p_1 p_2)}  \frac{d^3 \vec k_1 d^3 \vec k_2 d^3 \vec q_1d^3 \vec q_2}{ 16 k_1^0 k_2^0 q_1^0 q_2^0}
\end{eqnarray}
The delta-function of spatial momenta is taken off by integration over $q_2$.
To take off the delta-function over energy we rewrite
 $$ d^3 q_1 =  | \vec q_1|^2 d|\vec q_1| \sin \theta d\theta d\varphi $$ 
where $\theta$ and $\varphi$ is polar and axial angle relative to the vector $\vec k_1+\vec k_2-\vec p_1-\vec p_2$.
We also rewrite the argument of the delta-function as 
\begin{eqnarray}
 |\vec q_1| + \sqrt{|\vec q_1|^2 + |\vec k_1+\vec k_2 -\vec p_1 - \vec p_2|^2 + 2|\vec q_1| |\vec k_1+\vec k_2 -\vec p_1 - \vec p_2| \cos \theta}  +\nonumber\\
 |\vec k_1| + |\vec k_2| - |\vec p_1| - |\vec p_2| \nonumber 
\end{eqnarray}
After integration over $|\vec q_1|$ we obtain
\begin{eqnarray}
d \sigma =  \frac{4 g^8 (k_1 p_2) (q_2 k_2) ((k_1 p_2)(p_1 q_1) + (p_1(k_1-p_2))(q_1(k_1-p_2))       ) }{(2\pi)^8 (((p_1+p_2)^2-M_W^2)^2+M_W^2\Gamma_W^2) (((q_1+q_2)^2-M_W^2)^2+M_W^2 \Gamma_W^2 )     } \nonumber\\
\frac{|\vec q_1| d^3 \vec k_1 d^3\vec k_2 d\theta \sin \theta d \varphi}{16 (p_1 p_2) (((p_1+p_2-k_1)^2 - m_{\nu}^2)^2 + m_{\nu}^2 \Gamma_{\nu}^2)  |\vec k_1| |\vec k_2|  (|\vec q_1| + |\vec q_2| + |\vec q_1+\vec q_2|\cos \theta) }
\label{CSF-la}
\end{eqnarray}
Here we also introduced the $\Gamma_W$, $\Gamma_{\nu}$ which are the widths of decay of $W$ and $\nu$.
Let us also note that $|\vec k_1|$ and $|\vec k_2|$ does not produce any divergence because if any of them is close to zero, then the appropriate scalar product $(k_1 p_2)$ or $(q_2 k_2)$ is also close to zero.

The amplitude of the reaction represented in Figure \ref{MajoranaDiagram} is
equal to

\begin{eqnarray}
\bar{v}(p_2) \frac{ig}{\sqrt{2}} \gamma^{\mu} \frac{(1+\gamma^5)}{2} u(p_1) \frac{-i}{(p_1+p_2)^2 - M_W^2} \left( g_{\mu\nu} \frac{(p_1+p_2)_{\mu}(p_1+p_2)_{\nu}}{M_W^2} \right) \nonumber\\
\bar{u}(k_1) \frac{ig}{\sqrt{2}} \gamma^{\nu} \frac{(1+\gamma^5)}{2} \frac{i m_{\nu}C \gamma^0}{(p_1+p_2-k_1)^2 - m_{\nu}^2} \frac{ig}{\sqrt{2}} \gamma^{\lambda} \frac{(1+\gamma^5)}{2} u(k_2) \nonumber\\
\frac{-i}{(q_1+q_2)^2 - M_W^2}  \left( g_{\lambda \rho} \frac{(q_1+q_2)_{\lambda}(q_1+q_2)_{\rho}}{M_W^2} \right) \frac{ig}{\sqrt{2}} \bar{u} (q_2) \gamma^{\rho} \frac{(1+\gamma^5)}{2} v(q_1)
\end{eqnarray}
Its squared amplitude can be transformed to 
\begin{eqnarray}
|M|^2 = \frac{8 g^8 (k_1 p_2) (p_1 q_1) (q_2 k_2)  }{ ((p_1+p_2)^2-M_W^2)^2 ((q_1+q_2)^2-M_W^2) ((p_1+p_2-k_1)^2 - m_{\nu}^2)     }
\end{eqnarray}
(This answer is also obtained by using \textsc{mathematica}). In the same way one can show that the cross-section for this reaction is:
\begin{eqnarray}
d \sigma =  \frac{8 g^8 (k_1 p_2)  (p_1 q_1) (q_2 k_2)         }{(2\pi)^8 (((p_1+p_2)^2-M_W^2)^2+M_W^2\Gamma_W^2) (((q_1+q_2)^2-M_W^2)^2+M_W^2 \Gamma_W^2 )     } \nonumber\\
\frac{|\vec q_1| d^3 \vec k_1 d^3\vec k_2 d\theta \sin \theta d \varphi}{16 (p_1 p_2) (((p_1+p_2-k_1)^2 - m_{\nu}^2)^2 + m_{\nu}^2 \Gamma_{\nu}^2)  |\vec k_1| |\vec k_2|  (|\vec q_1| + |\vec q_2| + |\vec q_1+\vec q_2|\cos \theta) }
\label{CSF-la1}
\end{eqnarray}

\section{The process as a decay chain.}
\label{Decay}

\begin{figure}[h!]
\begin{center}
\includegraphics[width=16cm,angle=0]{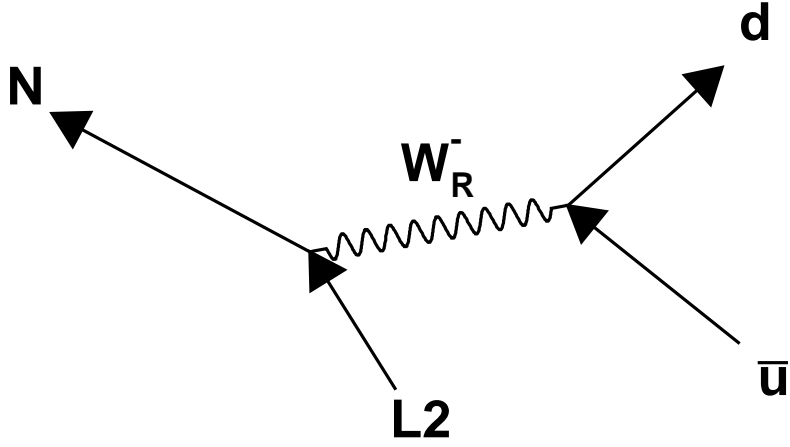}
\caption{The process represented as two consecutive decays}
\label{DecayDiagram}
\end{center}
\end{figure}

As we said in the introduction, instead of considering the whole process, we
can assume that after the production in the collision of the quark and
antiquark the $W^-_R$ boson decays into the lepton and heavy neutrino N
and then the heavy neutrino N decays into the lepton and the quark-antiquark
pair. The diagram in Figure \ref{DecayDiagram} shows the last part
of this process (the first part is trivial since we consider here the
production of $W_R$ at rest).


In this section we calculate the width of this decay (it does not depend on
the type of the neutrino). The expression for the amplitude is the following:
\begin{eqnarray}
iM = \bar v_N  \frac{ig}{\sqrt{2}} \gamma^{\mu} \frac{(1+\gamma^5)}{2} v_L \frac{-i}{(q_1+q_2)^2 - M_W^2}\nonumber\\
 \left( g_{\mu\nu} - \frac{(q_1+q_2)_{\mu}(q_1+q_2)_{\nu}}{M_W^2} \right)
  \bar u_d \frac{ig}{\sqrt{2}} \gamma^{\nu} \frac{(1+\gamma^5)}{2} v_u
 \end{eqnarray}
Its squared modul can be simplified to
\begin{eqnarray}
|M|^2 = \frac{2 g^4 (L q_1)(k_2 q_2)}{\left(( (q_1+q_2)^2 - M_W^2)^2 + M_W^2 \Gamma_W^2\right)},
\end{eqnarray}
where $L$ is the momentum of the neutrino and the other momenta are the same as in the previous section.
We substitute this formula into the general expression for the width of decay \cite{LL}
\begin{eqnarray}
dw = \frac{1}{(2\pi)^5} \delta^{(4)} (k_2+q_1+q_2-L) |M|^2 \frac{d^3 \vec k_2 d^3 \vec q_1 d^3 \vec q_2}{16 L^0 k_2^0 q_1^0 q_2^0} 
\end{eqnarray}
Again the delta-function of spatial momenta is taken off by integration over $q_2$. The delta-function of energy is also taken off in the same way. We rewrite
$$  d^3 \vec q_1 = |\vec q_1|^2 d |\vec q_1| \sin \theta d \theta d\varphi ,$$
where $\theta$ and $\varphi$ is polar and axial angle of $\vec q_1$ relative to the vector $\vec L - \vec k_2 $, and 
$$ L^0 - k_2^0 - q_1^0 -q_2^0  = \sqrt{\vec L^2 + M_{\nu}^2} - |\vec k_2| - |\vec q_1| - \sqrt{|\vec L-\vec k_2|^2 +  |\vec q_1|^2 - 2|\vec q_2||\vec L  - \vec k_2| \cos \theta}. $$
After integration over $|\vec q_1|$ we obtain that
\begin{eqnarray}
dw = \frac{|M|^2 |\vec q_1| \sin \theta d^3\vec k_2 d \theta d \varphi}{(2\pi)^5 16 (|q_1|+|q_2|+|q_1+q_2|\cos\theta) |k_2+q_1+q_2| |k_2|} \label{decay}
\end{eqnarray}

\section{Description of the programs}

We compare the angular distributions of the
final state particles obtained using the full and simplified calculations.
Two different programs perform the Monte-Carlo simulation of the process
using the von Neumann's method. The first program uses the cross-section
formula obtained in section \ref{Cross-section}, the second one uses
the simplified approach considered in section \ref{Decay}. More details
of this simulation follow.

In this estimation we assume that the masses of $W_R$ and heavy neutrino
are equal to 1000 GeV and 500 GeV correspondingly. These are typical
values for the searches of these new particles at the LHC. For the moment
we assume that the initial quarks have the same energy and
opposite momentum directions so that the $W_R$ is produced at rest:
\begin{eqnarray}
p1 = \{E , 0 , 0 , E\}, \nonumber\\
p2 = \{E , 0 , 0 , -E\}\label{p_1,2},
\end{eqnarray}
where $E$ is the initial energy of quarks.

The first program has two similar versions: the first one considers the case
of Dirac-like neutrino where only the reaction 1 can take place,
the considers considers the case of Majorana-like neutrino
where both reactions 1 and 2 happen with a probability of 50\% each.

The maximum of the cross section is determined (maximization step) by the
Monte Carlo method before the simulation step. At each iteration of the
corresponding loops the final state particles 4-momenta take random values
so that the energy and momentum are conserved. At the simulation step
2 or 4 loops are performed because antiquark can have positive or negative
$p_z$ and because reactions 1 and 2 are possible in the Majorana neutrino
case.

\section{Results}

\begin{figure}[p] 
\begin{center}
\includegraphics[width=16cm,angle=0]{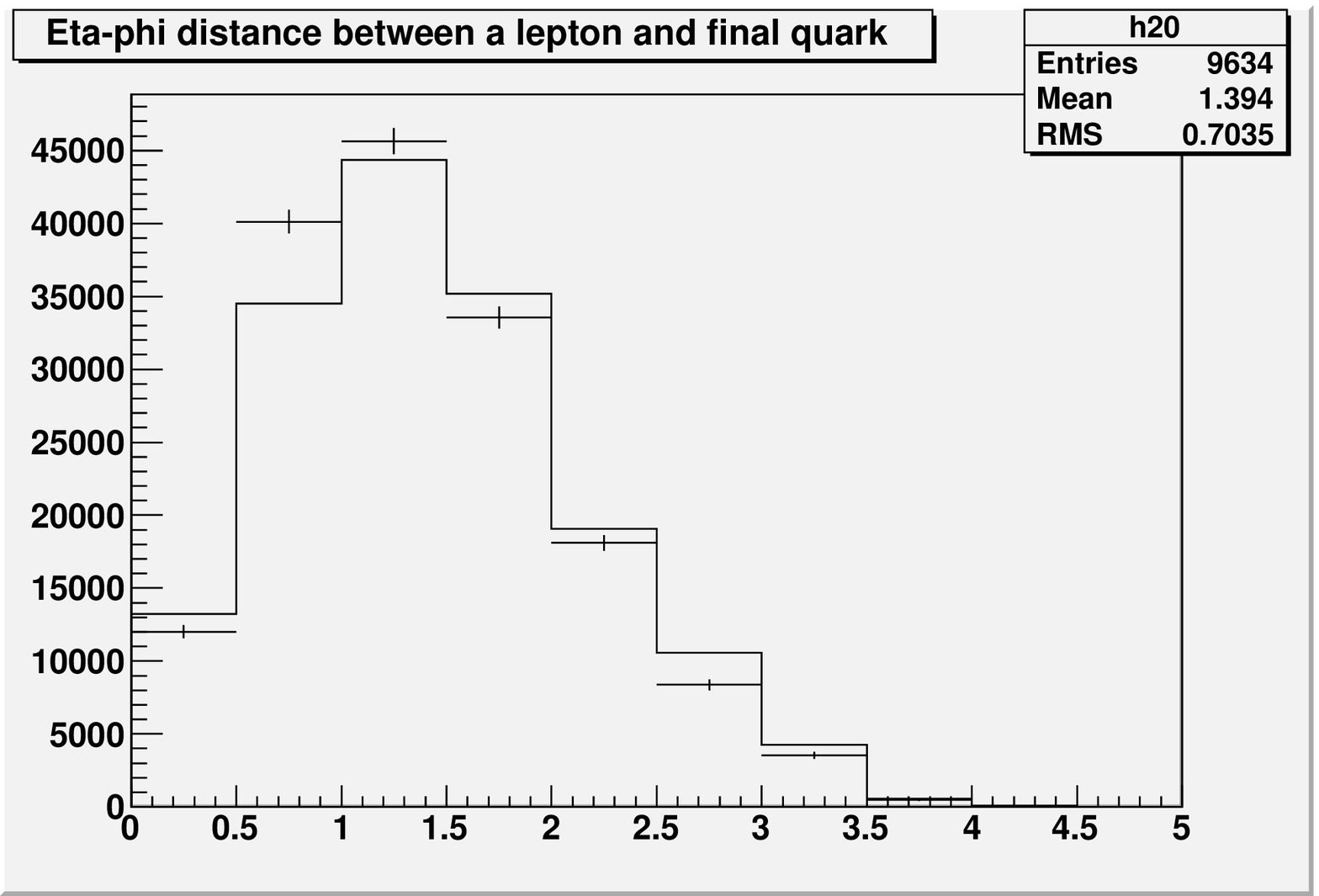}

\caption{Distributions of the minimal distance in the $\eta - \phi$ plane from
one of the final state leptons to one of the final state quarks. Histogram -
from the simulation as a decay chain (statistical errors are negligible),
points with errors - from the full matrix element with a Majorana heavy
neutrino.}
\label{Maj-decay}
\end{center}
\end{figure}

In Figure \ref{Maj-decay} one can see the results. They are represented as a
comparison of the distributions of the minimal distance $R$ in the
$\eta - \phi$ plane from
one of the final state leptons to one of the final state quarks. This
distribution is important because in real collider experiments the efficiency
of reconstruction drops when this distance is small (to exact zero when it
is smaller than 0.5). The shape is different. To estimate the systematic
error arising from the simulation as a decay chain we calculated the ratio
of the number of events with $R > 1$, where the drop of the lepton
reconstruction efficiency due to the presence of a nearby jet is small, to
the total number of events for the two simulations. They differ by 2.7\%.
This is our estimate of the systematic error of the simplified simulation.
The difference between the cases of Dirac and Majorana heavy neutrinos is
small (within 1\%).

\rule{0pt}{30mm}

\end{document}